\documentclass[11pt,a4paper]{article}
\pdfoutput=1
\usepackage{jheppub}

\usepackage{mathrsfs}
\usepackage{amsmath, amsthm, amssymb}
\usepackage{color}
\usepackage{amssymb}
\usepackage{verbatim}
\usepackage{multirow}
\usepackage[normalem]{ulem}

\newcommand{\be}{\begin{equation}}
\newcommand{\ee}{\end{equation}}
\newcommand{\bea}{\begin{eqnarray}}
\newcommand{\eea}{\end{eqnarray}}

\preprint{\begin{flushright} FTUAM-15-40 \\ IFT-UAM/CSIC-15-119\\ FERMILAB-PUB-15-484-T \end{flushright}}

\title{The MOMENT to search for CP violation}
\author[a]{Mattias Blennow,}
\emailAdd{emb@kth.se}
\affiliation[a]{\footnotesize Department of Theoretical Physics, School of Engineering Sciences, KTH Royal Institute of Technology, 
  Albanova University Center, 106 91 Stockholm, Sweden}

\author[b]{Pilar Coloma}
\emailAdd{pcoloma@fnal.gov}
\affiliation[b]{\footnotesize Theoretical Physics Department, Fermi National Accelerator Laboratory, \\
P.O. Box 500, Batavia, IL 60510, USA  }

\author[c,d]{and Enrique Fern\'andez-Mart\'inez}
\emailAdd{enrique.fernandez-martinez@uam.es}
\affiliation[c]{\footnotesize Departamento de F\'isica Te\'orica, Universidad Aut\'onoma de Madrid, Cantoblanco E-28049 Madrid, Spain}
\affiliation[d]{\footnotesize Instituto de F\'isica Te\'orica UAM/CSIC,
 Calle Nicol\'as Cabrera 13-15, Cantoblanco E-28049 Madrid, Spain}

\abstract{
In this letter, we analyze for the first time the physics reach in terms of sensitivity to leptonic CP violation of the proposed MuOn-decay MEdium baseline NeuTrino beam (MOMENT) experiment, a novel neutrino oscillation facility that would operate with neutrinos from muon decay. Apart from obtaining a sufficiently intense flux, the bottlenecks to the physics reach of this experiment will be achieving a high enough suppression of the atmospheric background and, particularly, attaining a sufficient level of charge identification. We thus present our results as a function of these two factors. As for the detector, we consider a very massive Gd-doped Water Cherenkov detector. We find that MOMENT will be competitive with other currently planned future oscillation experiments if a charge identification of at least 80~\% can be achieved at the same time that the atmospheric background can be suppressed by at least a factor of ten. We also find a large synergy of MOMENT with the current generation of neutrino oscillation experiments, T2K and NOvA, which significantly enhances its final sensitivity.
}

\begin{document}

\maketitle

\section{Introduction}

The violation of the charge-parity (CP) symmetry in Nature holds a very particular role in the development of modern theoretical physics. In the quark sector, the violation was observed in the decays of neutral kaons in 1964~\cite{Christenson:1964fg} and was fundamental in the prediction of the third generation of quarks~\cite{Kobayashi:1973fv}. Furthermore, CP-violation is also one of the Sakharov conditions~\cite{Sakharov:1967dj}, which describe the necessary ingredients for creating a baryon asymmetry in the early Universe. With the amount of CP-violation in the quark sector
being too small to account for the observed baryon asymmetry~\cite{Gavela:1993ts,Gavela:1994dt}, the discovery of a different source of CP-violation could prove crucial to further our understanding of the genesis of matter over anti-matter.

Possible additional sources of CP-violation can be found in the lepton sector, once the Standard Model (SM) is extended in order to include neutrino masses. The mixing of massive neutrinos in the flavor basis allows for 
the inclusion of non-trivial complex phases in the Pontecorvo--Maki--Nakagawa--Sakata (PMNS) matrix~\cite{Pontecorvo:1957cp,Pontecorvo:1957qd,Maki:1960ut,Maki:1962mu,Pontecorvo:1967fh}, in an analogous manner to what is done in the quark sector. Assuming that there are only three neutrino families the PMNS matrix will contain one or three such phases, depending on whether neutrinos are Dirac or Majorana particles. Although neutrino oscillation experiments are insensitive to the two Majorana CP-violating phases, they can probe the Dirac CP-violating phase. 

In the last few years, new results from the latest generation of neutrino oscillation experiments have started to provide precision measurements of the parameters describing the masses and mixing of neutrinos. In particular, with the measurements of the size of the PMNS matrix element $U_{e3}$ provided by accelerator and reactor neutrino experiments~\cite{An:2012eh,Ahn:2012nd,Abe:2012tg,Adamson:2011qu,Abe:2011sj}, it is plausible that CP-violation in the lepton sector may be found in the not so distant future. The current hints of maximal lepton CP-violation~\cite{Abe:2013hdq,NOvAFirstResults,Palazzo:2015gja} provide further indication that this discovery may be right around the corner.

In the next generation of neutrino oscillation experiments, the front runners in the hunt for leptonic CP-violation are the proposed Deep Underground Neutrino Experiment (DUNE)~\cite{DUNE_CDR} and the Tokai to Hyper-Kamiokande (T2HK) experiment~\cite{Abe:2011ts}. Both of them propose to use conventional accelerator neutrino beams from pion decay. In contrast, the MuOn-decay MEdium baseline NeuTrino beam facility (MOMENT)~\cite{Cao:2014bea} proposes to observe a neutrino beam produced from decaying muons at relatively low energies. By using this type of beam, some of the technical difficulties related to the construction of the more futuristic neutrino factory could be avoided~\cite{Geer:1997iz,DeRujula:1998hd,Bandyopadhyay:2007kx}. The aim of this letter is to study the capabilities of the MOMENT experiment and put it into context in the global experimental effort in neutrino physics.

\section{Implementation}

The MOMENT design is still not fully developed and is therefore subject to large uncertainties. As a first step towards studying its physics potential and the requirements it would need to meet to reach a competitive performance with respect to other future neutrino oscillation experiments, some assumptions regarding both the beam and detector performance have to be made. However, in our analysis we leave the most relevant parameters free in order to explore their impact on the expected sensitivities. 

The MOMENT facility would employ a proton linac (either continuous or pulsed) of 1.5 GeV, as well as a 10 mA proton driver. The aim of its design is to deliver a beam of extremely high power, up to 15 MW. Reaching such a high intensity already represents a major technological challenge. In addition, if such a high intensity is eventually achieved, a suitable target that is able to withstand it would need to be identified. Further issues have been pointed out related to the focusing system for the pions, heat mitigation and the radiation levels at the target station. These points are already being investigated, and we refer the interested reader to Ref.~\cite{Cao:2014bea}. In this work we will start from the muon and electron neutrino fluxes presented in Refs.~\cite{Cao:2014bea,yifanginvisibles} (at 150~km from the source), and we will assume that alternating between muon polarities with a similar flux intensity is possible. In order to assess the importance of achieving the demanding goal of 15~MW, we will also show how our results scale with the total luminosity of the experiment. The neutrino fluxes used in this work have their maximum at energies around 150~MeV with maximum intensity of $\sim 10^9$~MeV$^{-1}$~m$^{-2}$~year$^{-1}$, and have been taken from Ref.~\cite{yifanginvisibles}. Five years of running time per polarity are assumed. 

In principle, the MOMENT setup would allow the study of the $\nu_e \to \nu_e$, $\nu_\mu \to \nu_e$, $\nu_e \to \nu_\mu$ and $\nu_\mu \to \nu_\mu$ oscillation channels as well as their corresponding CP-conjugate partners. However, since the original flux is composed of $\nu_\mu$ and $\bar{\nu}_e$ from $\mu^-$ decay, both good flavour and charge identification capabilities are needed in order to be sensitive to a possible CP-violating signal. The neutrino flux for this facility would peak at low energies around 150-200~MeV. Therefore, a very massive detector would be required in order to compensate the low interaction cross section at these energies and reach large enough statistics. The detector technology for MOMENT has not yet been decided, but a massive Water Cherenkov detector has been suggested due to its excellent flavour identification capabilities and performance at low energies. The drawback of using a Water Cherenkov in combination with the MOMENT beam is its inability to distinguish neutrinos and antineutrinos. Nevertheless, this problem may be solved (at least partially) by doping the water with Gd~\cite{Beacom:2003nk} at the 0.1-0.2\% level. We will thus adopt a Mton class (500~kton fiducial) Gd-doped Water Cherenkov detector as baseline detector for our analysis. 

In this study, the detector response has been implemented following Ref.~\cite{Agostino:2012fd}. Migration matrices, describing both the detection efficiencies and energy reconstruction, are used for all four relevant oscillation channels (and their CP-conjugates). The most relevant backgrounds come from charge mis-identification (charge mis-ID) of events coming from the intrinsic contamination of the beam, flavour mis-identification and neutral current (NC) backgrounds mis-identified as charged current (CC) events. Since charge mis-ID will be one of the bottlenecks for the physics performance of the facility, our results will be presented as a function of this parameter. In Ref.~\cite{Huber:2008yx} it was estimated that Gd-doping alone (at the 0.1-0.2~\% level) could bring charge separation up to the 80~\% level. Besides Gd-doping, some statistical neutrino/antineutrino discrimination could be achieved from other distinctive features~\cite{Huber:2008yx}, such as the angular distribution between the charged lepton and the incident neutrino/antineutrino, or the different lifetimes of the outgoing muons/antimuons produced in $\nu_\mu$/$\bar\nu_\mu$ interactions. Since it is uncertain how much extra charge-identification efficiency these extra handles would eventually bring to the table\footnote{Very recently, Gd-doping has been approved for the Super-KamiokaNDE detector~\cite{gadzookstalk2}. Therefore, by the time a Mton-class Water Cherenkov detector is built, the behavior of this detector technology will be well understood. }, we will show how much the performance of the setup would improve if the total charge-identification efficiency surpasses the 70~\% level, which is taken as a (conservative) lower threshold~\cite{markvagins}.

Another important limiting factor could be the potentially large atmospheric-induced background. By placing the detector deep underground all such background, except the contribution from atmospheric neutrinos, can be efficiently suppressed: at a depth of 2500 m of water equivalent, the muon flux would be reduced by almost two and a half orders of magnitude (see, e.g., Fig.3 in Ref.~\cite{Gray:2010nc}). We will therefore consider the background coming from particles interacting in the atmosphere to be negligible, with the sole exception of that coming from atmospheric neutrinos. This contribution, on the other hand, could be largely reduced by sending the neutrino flux in short bunches, so that a time cut can be efficiently applied. This is usually parametrized in terms of a \emph{suppression factor} (SF), \textit{i.e.,} the ratio between the length of each bunch to the distance between bunches. In neutrino oscillation experiments using pion decay beams, the achieved SF is typically around $10^{-3}$~\cite{Bandyopadhyay:2007kx}. In the current work, we will explicitly consider the atmospheric background, computed as in Ref.~\cite{FernandezMartinez:2009hb}, applying a SF ranging from 1 to $5 \cdot 10^{-3}$ in order to quantify its impact on the final sensitivities.  Finally, we also include an overall 5~\% (10~\%) normalization systematic error, uncorrelated between all signal (background) channels. All of our numerical simulations have been implemented using the GLoBES software~\cite{Huber:2004ka,Huber:2007ji}.

For convenience, Tab.~\ref{tab:events} summarizes the total expected event rates in the energy range between 0 and 1.6~GeV, for all oscillation channels under consideration, after efficiencies are accounted for. The signal and background rates are provided separately for each channel, assuming a charge separation efficiency of 70~\% and a suppression factor $\mathrm{SF}=10^{-1}$ for the atmospheric neutrino background. These number of events have been obtained assuming that the true values of the oscillation parameters correspond to the best-fit values from Ref.~\cite{Gonzalez-Garcia:2014bfa}, with the sole exception of the CP-violating phase which is set to $\delta = 0$. A normal ordering of the neutrino masses ($m_1 <m_2 <m_3$) has also been assumed. Only those events with reconstructed neutrino energy between 0.1 and 1~GeV are considered for the $\chi^2$ analysis. 

\begin{table}
\begin{center}
\renewcommand{\arraystretch}{1.6}
\begin{tabular}{r@{\quad} | c@{\quad} | c@{\quad} c@{\quad} c@{\quad} c@{\quad} }
Channel & Signal & NC & CID & FID & Atm. \\ \hline
$\nu_e \rightarrow \nu_\mu $ & 822 & 60 & 1004 & 11 & 652 \\
$\bar\nu_e \rightarrow \bar\nu_\mu $ & 292 & 212 & 2851 & 4 & 449 \\
$\nu_\mu \rightarrow \nu_e $ & 1044 & 41 & 3191 & 9 & 399 \\
$\bar\nu_\mu \rightarrow \bar\nu_e $ & 358 & 66 & 7567 & 4 & 268 \\
$\nu_\mu \rightarrow \nu_\mu $ & 6653 & 91 & 124 & 2 & 652 \\
$\bar\nu_\mu \rightarrow \bar\nu_\mu $ & 2343 & 138 & 352 & 5 & 449 \\
$\nu_e \rightarrow \nu_e $ & 17657 & 28 & 153 & 2 & 399 \\
$\bar\nu_e \rightarrow \bar\nu_e $ & 7445 & 96 & 448 & 4 & 268 \\ \hline
\end{tabular}
\caption{\label{tab:events} Total number of events (after oscillations) for all oscillation channels considered in the analysis. The number of signal and background events are given separately. Background contributions from neutral-current (NC), charge mis-identification (CID), flavor mis-identification (FID) and atmospheric (Atm) events are shown separately. A charge separation of 70~\% and a suppression factor $\mathrm{SF}=10^{-1}$ have been assumed.  }
\end{center}
\end{table}

\section{Results and conclusions}

In its most conservative incarnation, with a 70~\% charge ID and no suppression of the atmospheric background, we find that the MOMENT facility, on its own, barely improves over what the presently running experiments T2K and NO$\nu$A will achieve in the coming years. However, we have found that combining the data from the three facilities can be quite complementary, leading to a significant improvement of their individual physics reaches beyond that due to a simple increase in statistics. In the following, we have simulated the sensitivity from the NO$\nu$A experiment as in Ref.~\cite{Blennow:2013oma}, using 3 years of data taking per polarity and $6.0\times 10^{20}$~protons on target (PoT) per year. This is then combined with a simulation of T2K data using neutrino data corresponding to approximately $3\times 10^{20}$ PoT. The T2K fluxes have been taken from Ref.~\cite{Abe:2012av} and the signal and background efficiencies have been set to approximately match the results from Ref.~\cite{Abe:2015awa} for the same exposure.

%%%%%%%%%%%%%%%%%%%%%%%%%%%%%%%%%%%%%%%%%%%%%%
\begin{figure*}
\begin{center}
\includegraphics[width=1\textwidth]{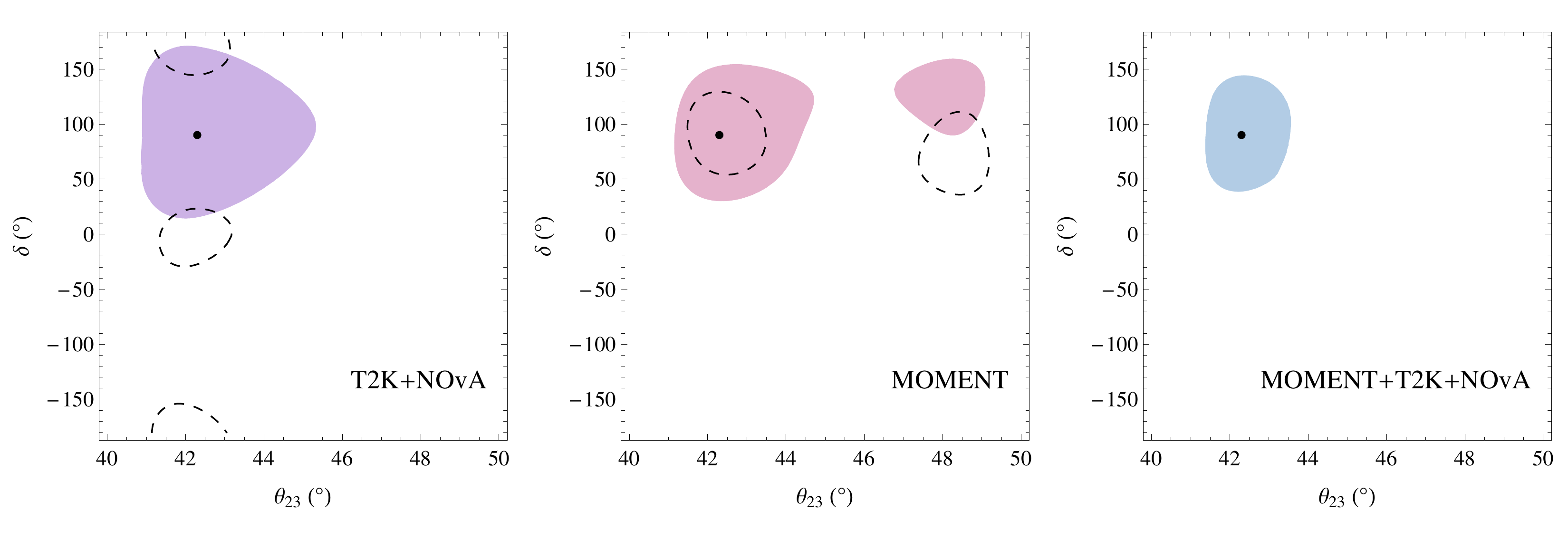}
\caption{\label{fig:pot} The complementarity between measurements at MOMENT and the data expected from the presently running facilities T2K and NO$\nu$A. In each panel, the shaded areas indicate the allowed confidence regions when the fit is done using the correct mass ordering (normal ordering, in this example), while the dashed lines indicate the allowed regions when the fit is performed using the wrong mass ordering (sign degeneracies). All regions correspond to 90~\% confidence level, for 2 d.o.f.. The black dot indicates the assumed true values for $\theta_{23}$ and $\delta$. }
\end{center}
\end{figure*}
%%%%%%%%%%%%%%%%%%%%%%%%%%%%%%%%%%%%%%%%%%%%%%

The complementarity between MOMENT and the current generation of neutrino oscillation experiments is shown for a particular point in the $\theta_{23}$-$\delta$ parameter space in Fig.~\ref{fig:pot}. In each panel, the shaded areas show the confidence regions obtained in the $\theta_{23}-\delta$ plane for the correct neutrino mass ordering, while the dashed lines show the allowed regions for the opposite mass ordering (\textit{a.k.a.}, sign degeneracies~\cite{Minakata:2001qm}). Each panel corresponds to the expected results for a given  facility (or combination thereof), as indicated in the legend. As can be seen from a comparison between the left and central panels, the sign degeneracies affect both the T2K+NO$\nu$A and the MOMENT setup, but appear at completely different values of $\delta$ due to the much weaker matter effects that characterize the latter. Furthermore, the octant degeneracy also plays an important role at MOMENT, while it is solved at T2K+NO$\nu$A (for this particular point in parameter space).

We found that, even though both the $\bar\nu_e \to \bar\nu_\mu$ and $\nu_\mu \to \nu_e$ channels are available at MOMENT for $\mu^-$ running, the former channel dominates the physics reach unless very optimistic charge ID is assumed. This can be understood as follows. On one hand, the ``wrong sign'' electrons from $\bar\nu_e$ disappearance completely overwhelm the signal in the $\nu_\mu \rightarrow \nu_e$ channel. On the other hand, the $\nu_\mu$ present in the beam are less of an issue for the $\bar\nu_e \to \bar\nu_\mu$ channel, since most them have already oscillated to $\nu_\tau$ when they reach the detector and therefore do not contribute to the muon-like CC sample. Thus, the physics reach from MOMENT and T2K+NO$\nu$A is dominated by different and complementary channels. The right panel of Fig.~\ref{fig:pot} shows how the combination of the three facilities is able to solve all degeneracies unambiguously and determine the correct value of $\theta_{23}$ and $\delta$ with an allowed region which is significantly reduced compared to the individual fits. 

Since by the time the MOMENT facility is built the T2K and NO$\nu$A facilities will have already finished taking data, we will present our results for the combination of MOMENT+T2K+NOvA only. Notice that this essentially improves the overall performance for the most conservative choices for the charge ID and SF of MOMENT, while it has little impact in the optimistic scenarios. Similarly, the physics reach of DUNE or T2HK is mildly affected after combination with T2K+NO$\nu$A, since their observations are less complementary and do not lead to further degeneracy solving besides a small increase in statistics. For this reason, when comparing the reach of MOMENT to that of T2HK or DUNE, we will take the expected physics reach for the latter from their respective proposals. 

We have also explored the effect of changing the baseline of the MOMENT detector. Indeed, it has been shown that, given the relatively large value of $\theta_{13}$, if the neutrino flux is centered around the second oscillation peak, the sensitivity to $\delta$~\cite{Coloma:2011pg,Marciano:2001tz} improves considerably. This has been studied in depth for a similar low-energy neutrino beam, the ESS$\nu$SB~\cite{Baussan:2013zcy}, also in combination with a Water Cherenkov detector. In the case of MOMENT, we find that when the most conservative assumptions are made, the optimal baseline is around $L=150$~km. However, when the most optimistic assumptions are adopted, the sensitivity becomes almost independent of the baseline as it is increased from the first to the second peak. This is mainly due to the strong dependence on $\delta$ at longer baselines, which compensates for the lower statistics. Thus, in the following we will only consider a $L=150$~km baseline, since the performance of the detector is still uncertain.

%%%%%%%%%%%%%%%%%%%%%%%%%%%%%%%%%%%%%%%%%%%%%%
\begin{figure*}
\begin{center}
\includegraphics[width=0.35\textwidth]{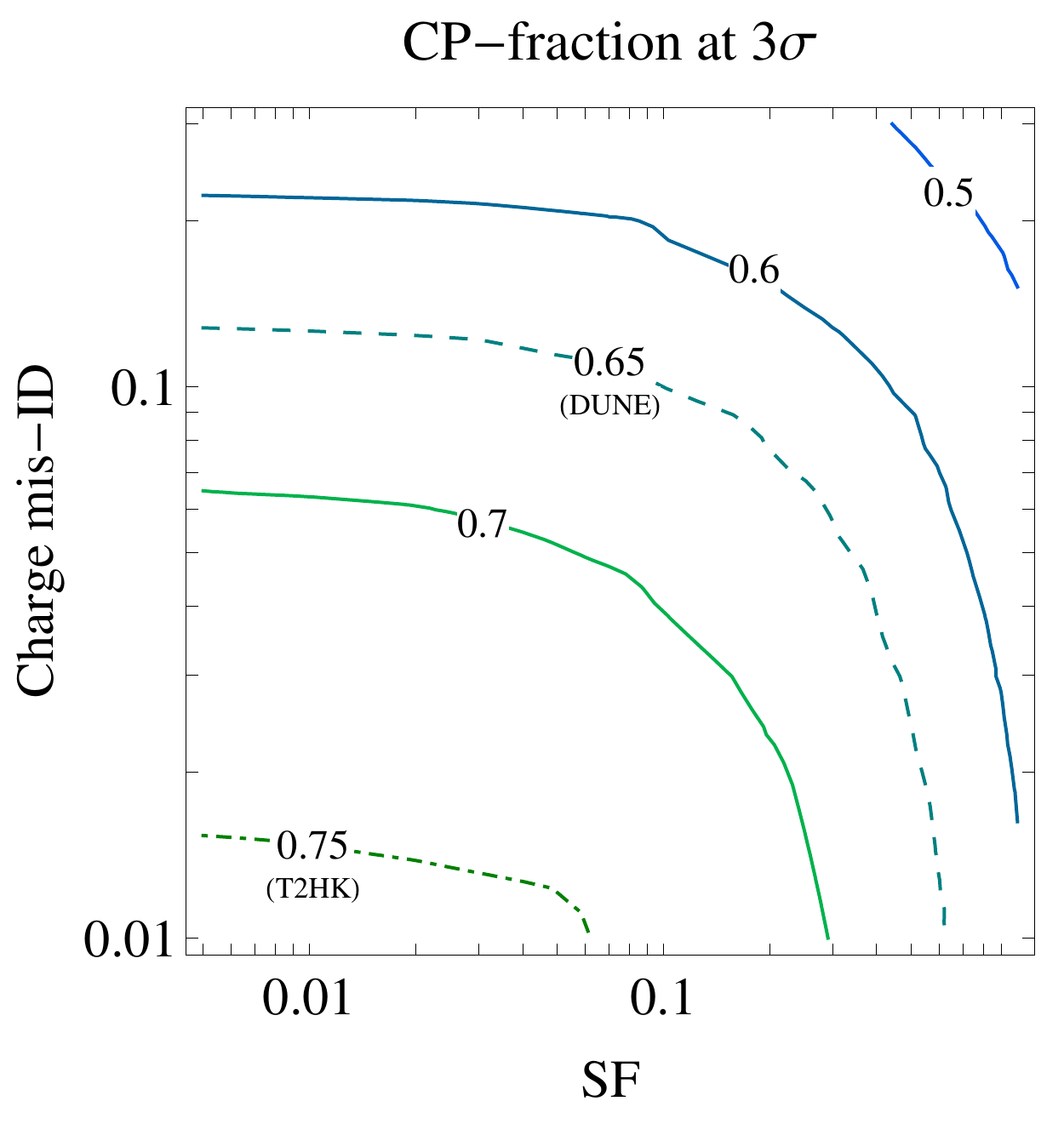} 
\includegraphics[width=0.35\textwidth]{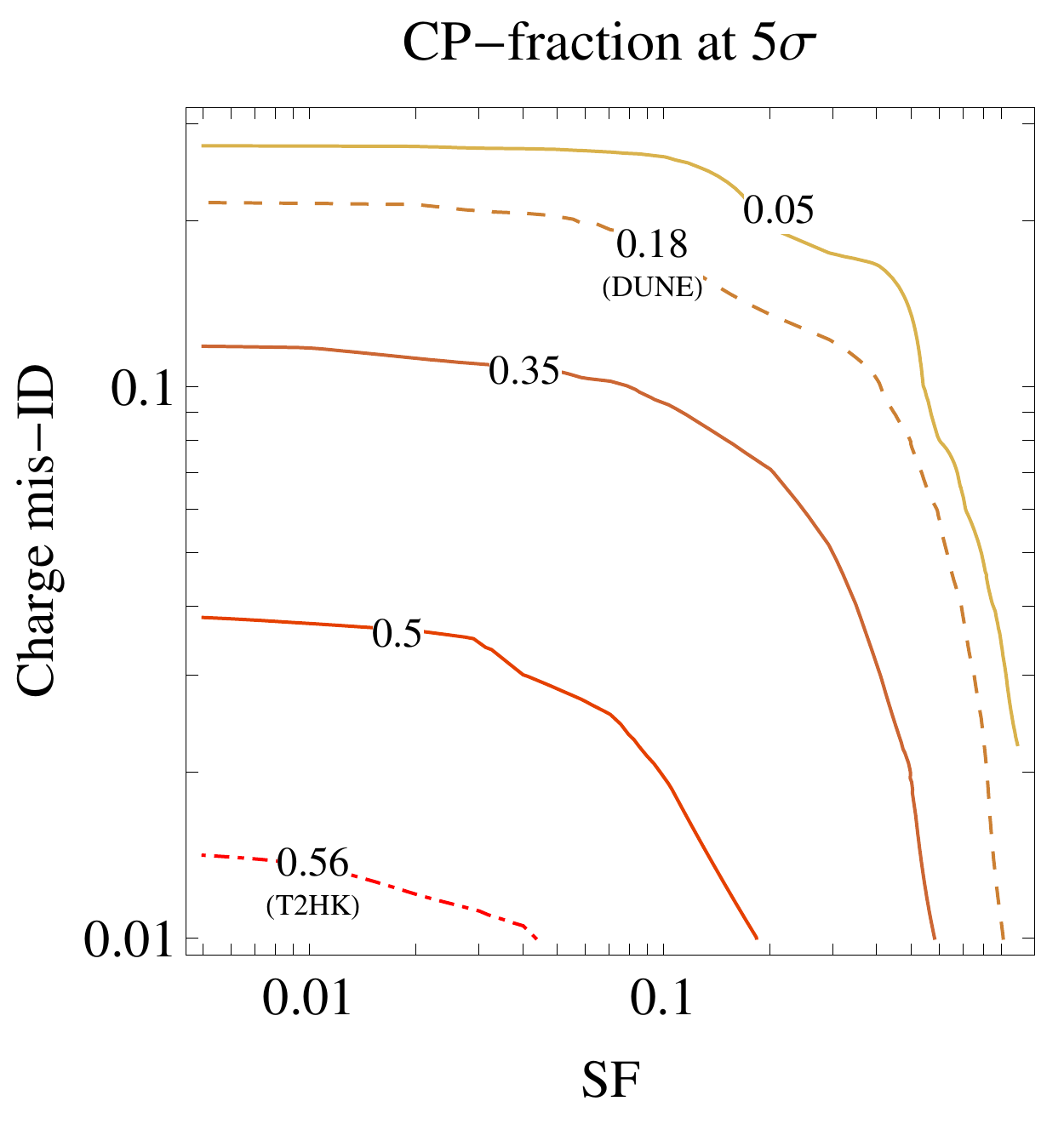}
\caption{\label{fig:sens} Fraction of values of $\delta$ for which a 3 (left panel) or 5 (right panel) $\sigma$ discovery of CP violation would be possible for the combination of MOMENT+T2K+NO$\nu$A, as a function of the achievable atmospheric background suppression factor (SF) and charge mis-identification rate (charge mis-ID) at the detector. In the region to the left/bottom of each line, the CP-fraction would be larger than the value indicated in each case. The dashed lines indicate the approximate reach for the DUNE experiment in each case (taken from Ref.~\cite{DUNE_CDR}), while the T2HK reach is indicated by the dot-dashed lines (taken from Ref.~\cite{Abe:2015zbg}).}
\end{center}
\end{figure*}
%%%%%%%%%%%%%%%%%%%%%%%%%%%%%%%%%%%%%%%%%%%%%%

Our main results are shown in Fig.~\ref{fig:sens}, where we show the fraction of possible values of $\delta$ for which the combination of MOMENT+T2K+NO$\nu$A would allow a $3 \sigma$ ($5 \sigma$) discovery of leptonic CP violation. Our results are shown as a function of the achievable charge-ID and atmospheric suppression factor. As can be seen, if a $\sim 80~\%$ charge-ID can be achieved, a 3 (5)~$\sigma$ discovery of CP violation would be possible for roughly $60~\%$ ($20~\%$) of the values of $\delta$, as long as the atmospheric suppression factor remains below $\mathrm{SF} \lesssim 0.1$. This is similar to the sensitivity reach expected for DUNE~\cite{DUNE_CDR} with an exposure of 300~MW$\cdot$kt$\cdot$yr  (corresponding to $\sim 3.5$ years running per polarity). Conversely, if the charge identification cannot be improved beyond $\sim 70~\%$, less than $10~\%$ of the values of $\delta$ would lead to a $5 \sigma$ discovery regardless of the value of SF. As a comparison, T2HK~\cite{Abe:2015zbg} with a 10-year run using a beam power of 750~MW would allow to cover $\sim 75~\%$ ($\sim 55~\%$) of the values of $\delta$ for a 3 ($5 \sigma$) discovery. MOMENT would require a charge ID of $\sim 98~\%$ and $\mathrm{SF}< 5 \cdot 10^{-2}$ to achieve a similar performance.
 
\begin{figure*}
\begin{center}
\includegraphics[width=0.65\textwidth]{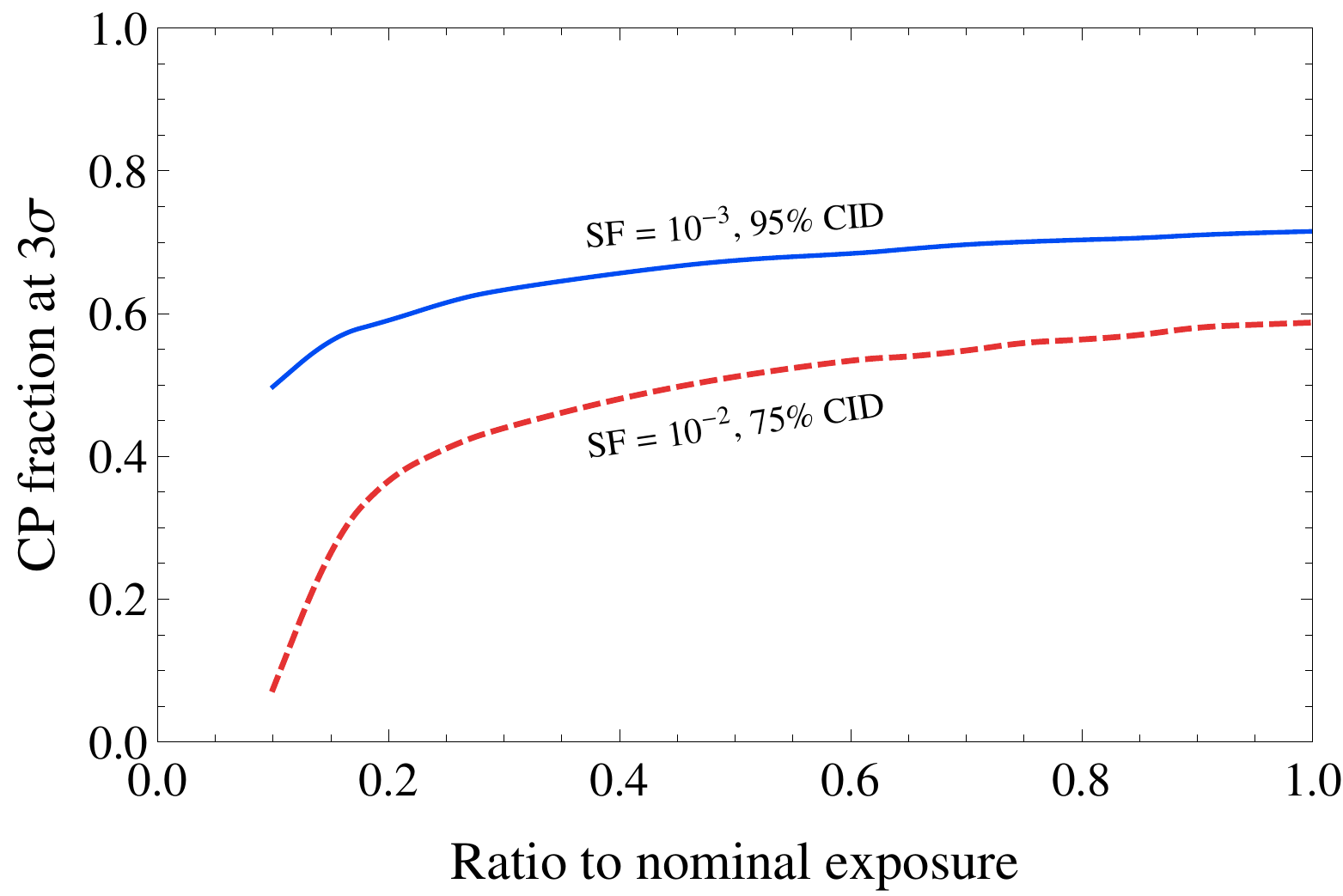}
\caption{\label{fig:expo} Fraction of values of $\delta$ for which a 3$\sigma$ discovery of CP violation would be obtained, as a function of the ratio of the considered exposure to the nominal exposure considered in this work ($500$~kt$\times 15$~MW$\times 10$~yr). }
\end{center}
\end{figure*}

Finally, as it was already mentioned, the MOMENT beam will have several technical challenges to meet before reaching its nominal beam intensity. Therefore, we have also studied the impact of the total exposure on the performance of the facility. This is shown in Fig.~\ref{fig:expo}, where we show the fraction of values of $\delta$ for which CP violation could be observed at the 3$\sigma$ level, as a function of the ratio between the considered exposure to the nominal exposure. Results are shown under two different sets of assumptions, as indicated in the figure, for the suppression factor and charge-identification capability of the detector. As it can be seen from the figure, the performance of the facility is not limited by statistics and therefore the total exposure can be reduced by a factor of between 5 and 10 before seeing a noticeable reduction in performance. This is due to the fact that most of the background is beam-related, and therefore the signal to background ratio does not change much when the exposure is reduced. At some point the atmospheric background dominates over the beam-induced and the decrease in sensitivity becomes much more pronounced. This situation is reached earlier for the more conservative assumption as expected as can be seen in the figure. A qualitatively similar behavior is also found at higher confidence levels, although the decrease in the CP coverage takes place sooner as the exposure of the experiment is decreased (as expected).

%%%%%%%%%%%%%%%%%%%%%%%%%%%%%%%%%%%%%%%%%%%%%%
%%%%%%%%%%%%%%%%%%%%%%%%%%%%%%%%%%%%%%%%%%%%%%

In conclusion, we have studied for the first time the physics reach attainable at MOMENT in terms of its CP violation discovery potential. We find that the main limiting factors to its performance are the charge identification and atmospheric background suppression. With a conservative assumption of $70~\%$ charge identification and no atmospheric background suppression, MOMENT would not improve significantly over the results expected at the end of the running period of T2K and NO$\nu$A, even after a 10 year run with a Mton Water Cherenkov detector. However, its combination with present facilities is able to lift several degeneracies and significantly improve the combined physics reach over a simple addition of statistics.  

In order to compete with other future neutrino oscillation facilities, more demanding detection capabilities would be necessary. We find that the physics reach of MOMENT would be similar to a 7 year run of DUNE if a charge identification of $\sim 80~\%$ and atmospheric suppression by a factor of 10 is achieved. To compete with 10 years of T2HK with a 750~MW beam, the background suppression factor should improve by a factor 20 keeping charge identification capabilities at the level of $\sim 98~\%$. In order to satisfy this requirement, a different detector technology would most likely be required in this case. 

\textit{\textbf{Acknowledgments.}} The work of MB was supported by the G\"oran Gustafsson Foundation. PC and EFM acknowledge financial support from the European Union through the ITN INVISIBLES (Marie Curie Actions, PITN-GA-2011-289442-INVISIBLES). EFM also acknowledges support from the EU through the FP7 Marie Curie Actions CIG NeuProbes (PCIG11-GA-2012-321582) and the Spanish MINECO through the “Ramon y Cajal” programme (RYC2011-07710) and the project FPA2009-09017. MB and EFM were also supported by the Spanish MINECO through the Centro de excelencia Severo Ochoa Program under grant SEV-2012-0249.
MB would also like to thank the Instituto de F\'isica Te\'orica, Madrid, for warm hospitality during the time when this work was initiated.
PC would like to thank the Mainz Institute for Theoretical Physics for hospitality and partial support during completion of this work. Fermilab is operated by Fermi Research Alliance, LLC under Contract No. \protect{DE-AC02-07CH11359} with the United States Department of Energy.

%\bibliographystyle{JHEP}
%\bibliography{refs}

\providecommand{\href}[2]{#2}\begingroup\raggedright\endgroup

\end{document}